\documentclass[usenatbib]{mn2e}
\usepackage{mathtools}
\usepackage{graphicx}
\usepackage{epstopdf} 
\usepackage{natbib}
\usepackage{verbatim}

\newcommand{\ciiid}{\hbox{[C\,{\sc iii}]$\lambda1907$+C\,{\sc iii}]$\lambda1909$}}

\newcommand{\Us}{\hbox{$U_\mathrm{S}$}}
\newcommand{\xid}{\hbox{$\xi_\mathrm{d}$}}
\newcommand{\nH}{\hbox{$n_{\mathrm{H}}$}}
\newcommand{\tauSFR}{\hbox{$\tau_\mathrm{SFR}$}}
\newcommand{\zform}{\hbox{$z_\mathrm{form}$}}
\newcommand{\tauV}{\hbox{$\hat{\tau}_V$}}
\newcommand{\Hii}{\mbox{H\,{\sc ii}}}
\newcommand{\subISM}{\textnormal{\tiny \textsc{ism}}}
\newcommand{\zism}{\hbox{$Z_\subISM$}}

 \usepackage{times}
\renewcommand{\thefootnote}{\fnsymbol{footnote}}

\def\lsim{\mathrel{\rlap{\lower4pt\hbox{\hskip1pt$\sim$}}
    \raise1pt\hbox{$<$}}}                
\def\gsim{\mathrel{\rlap{\lower4pt\hbox{\hskip1pt$\sim$}}
    \raise1pt\hbox{$>$}}}                

\begin{document}

\title[]
{Ly$\alpha$ and CIII] Emission in $z=7-9$ Galaxies: Accelerated Reionization Around Luminous Star Forming Systems?}

\author[Stark et al.] 
{Daniel P. Stark$^{1}$\footnotemark[1], 
Richard S. Ellis$^{2,3}$,
St\'{e}phane Charlot$^{4}$,
Jacopo Chevallard$^{5}$,
Mengtao Tang$^1$,
\newauthor
Sirio Belli$^{6}$,
Adi Zitrin$^{7}$\footnotemark[2],
Ramesh Mainali$^{1}$, 
Julia Gutkin,$^{4}$
Alba Vidal-Garc\'ia,$^{4}$ \newauthor
Rychard Bouwens $^{8}$, \&
Pascal Oesch$^9$
\vspace{0.1in}\\
$^{1}$ Steward Observatory, University of Arizona, 933 N Cherry Ave, Tucson, AZ 85721 USA \\  
$^{2}$ European  Southern  Observatory  (ESO), Karl-Schwarzschild-Strasse 2, 85748 Garching, Germany\\
$^{3}$ Department of Physics and Astronomy, University College London, Gower Street, London, WC1E 6BT, UK \\
$^{4}$ Sorbonne Universit\'{e}s, UPMC-CNRS, UMR7095, Institut d'Astrophysique de Paris, F-75014 Paris, France \\
$^{5}$ Scientific Support Office, Directorate of Science and Robotic Exploration, ESA/ESTEC, Keplerlaan 1, 2201 AZ Noordwijk, The Netherlands \\
$^{6}$  Max-Planck-Institut fur extraterrestrische Physik, Giessen-bachstr. 1, D-85741 Garching, Germany \\
$^{7}$ Cahill Center for Astronomy and Astrophysics, California Institute of Technology, MC 249-17, Pasadena, CA 91125, USA \\
$^{8}$  Leiden Observatory, Leiden University, NL-2300 RA Leiden, the Netherlands  \\
$^{9}$  Yale Center for Astronomy and Astrophysics, Department of Astronomy, Yale University, USA}

\pagerange{\pageref{firstpage}--\pageref{lastpage}} \pubyear{2016}

\hsize=6truein
\maketitle

\label{firstpage}
\begin{abstract}

We discuss new Keck/MOSFIRE spectroscopic observations of four luminous galaxies at 
$z\simeq 7-9$ selected to have intense rest-frame optical line emission by \citet{Roberts-borsani2015}.  Previous spectroscopic 
follow-up has revealed Ly$\alpha$ emission in two of the four galaxies.  Our new MOSFIRE observations confirm that
Ly$\alpha$ is present in the entire sample.   We detect  Ly$\alpha$ emission in the galaxy COS-zs7-1, confirming its redshift as 
$z_{\rm{Ly\alpha}}=7.154$, and we detect Ly$\alpha$ in EGS-zs8-2 at $z_{\rm{Ly\alpha}}=7.477$, verifying a tentative detection 
presented in an earlier study.  The ubiquity of Ly$\alpha$ emission in this unique photometric sample is 
puzzling given that the IGM is expected to be significantly neutral over $7<z<9$.   To investigate this surprising result 
in more detail, we have initiated a campaign to target UV metal line emission in the four Ly$\alpha$ emitters as a probe of both 
the ionizing radiation field and the velocity offset of Ly$\alpha$ at early times.
Here we present the detection of very large equivalent width [CIII], CIII] $\lambda\lambda$1907,1909 \AA\ 
emission in EGS-zs8-1 (W$_{\rm{CIII],0}}=22\pm2$~\AA), a galaxy from this sample previously shown to have Ly$\alpha$ 
emission at $z=7.73$. Photoionization models indicate that an intense radiation field (log$_{\rm{10}}$ $\xi^{\ast}_{ion}$ [erg$^{-1}$ Hz]$~\simeq 25.6$) and moderately low 
metallicity (0.11 Z$_\odot$) are required to reproduce the CIII] line emission and intense optical
line emission implied by the broadband SED. We argue that this extreme radiation field is likely to affect the local environment, increasing the transmission 
of Ly$\alpha$ through the galaxy.  Moreover, the centroid of CIII] emission indicates that Ly$\alpha$ is redshifted from the systemic
value by 340 km sec$^{-1}$. This velocity offset is larger than that seen in less luminous systems and provides an additional explanation 
for the transmission of Ly$\alpha$ emission through the intergalactic medium.  Since the transmission is further enhanced by the likelihood
that such systems  are also situated in the densest regions with accelerated evolution and the largest ionized bubbles, the visibility of Ly$\alpha$ at $z>7$ is expected 
to be strongly luminosity-dependent, with the most effective transmission occurring in systems with intense star formation.

\end{abstract}


\renewcommand{\thefootnote}{\fnsymbol{footnote}}
\footnotetext[1]{E-mail:dpstark@email.arizona.edu}
\footnotetext[2]{Hubble Fellow.}

\section{Introduction}

The reionization of intergalactic hydrogen is an important milestone in early cosmic history, marking the point 
at which nearly every baryon in the universe was affected by the growth of structure. 
How and when reionization occurs encodes unique insight into the nature of the first luminous 
objects, motivating a number of dedicated observational efforts aimed at studying the process.
Significant progress has been made in the last decade.   Measurement of the 
optical depth to electron scattering faced by the CMB reveals  
that the process is underway by $z\simeq 9$ \citep{Planck2015,Planck2016}, while 
quasar absorption spectra indicate that reionization is largely complete by $z\simeq 6$ 
(e.g., \citealt{Fan2006,Mcgreer2015}).   The large abundance of faint $z\gsim 6$ galaxies identified 
photometrically with the WFC3/IR camera on the {\it Hubble Space Telescope} (e.g., \citealt{McLure2013,Bouwens2015, Finkelstein2015}) 
suggests that the ionizing output of star forming systems may be sufficient to complete reionization by $z\simeq 6$ while also supplying the 
IGM with enough free electrons at $z\simeq 9$ to reproduce  the measured Thomson scattering optical depth of the CMB 
\citep{Robertson2015,Bouwens2015c,Stanway2016}.

New insight  is now being provided by spectroscopic surveys targeting Ly$\alpha$ emission from star forming galaxies at 
$z\gsim 6$.  Since Ly$\alpha$ is resonantly scattered by neutral hydrogen, the fraction of galaxies that
exhibit prominent Ly$\alpha$ emission should fall abruptly during the reionization era \citep{Fontana2010,Stark2010}.   Throughout the past five years, 
a large investment has been devoted to  searches for Ly$\alpha$ emission in the reionization era, resulting in only nine robust 
detections of Ly$\alpha$ at $z>7$ \citep{Vanzella2011,Ono2012, Schenker2012,Shibuya2012,Finkelstein2013,Oesch2015,Zitrin2015,Song2016}.    
These surveys clearly reveal a rapidly declining Ly$\alpha$ emitter fraction over $6<z<8$  (e.g., \citealt{Stark2010,Fontana2010,Ono2012,Pentericci2014,Tilvi2014,Schenker2014}), 
similar to the drop in the abundance 
of narrowband-selected Ly$\alpha$ emitters over $5.7<z<7.3$ (e.g. \citealt{Konno2014}).  
The decline in the volume density of  Ly$\alpha$ emitters at $z>6.5$  is consistent with strong attenuation 
from intergalactic hydrogen, possibly requiring neutral  fractions of x$_{\rm{HI}}\gsim 0.3-0.5$ at $z\simeq 7-8$ 
\citep{Mesinger2014,Choudhury2015}.  In this framework, the  small sample of known $z>7$ systems with 
detectable Ly$\alpha$ emission is thought to be galaxies that are situated in the largest ionized regions of the IGM, 
allowing  Ly$\alpha$ to redshift well into the damping wing before encountering intergalactic hydrogen. 

The first clues that this physical picture may be incomplete have recently begun to emerge.  
The  detection of nebular CIV emission in a low luminosity  $z=7.045$ Ly$\alpha$ emitter (A1703-zd6) led 
\citet{Stark2015b} to speculate that this galaxy's hard ionizing spectrum may  enhance its Ly$\alpha$ transmission by efficiently 
ionizing  surrounding hydrogen.   If true, this would suggest that observed counts of $z>7$ Ly$\alpha$ emitters may also 
depend on the prevalence of galaxies with extreme radiation fields, adding  uncertainty to the 
modeling of the evolving Ly$\alpha$ transmission at $z>6$.   Observational efforts are now underway to establish how 
common such hard ionizing spectra are among reionization-era galaxies and to constrain the powering mechanism 
(AGN or metal poor stars) of the nebular CIV emission (e.g., \citealt{Feltre2016}).

Perhaps even more puzzling is the discovery of Ly$\alpha$  in the first two galaxies observed 
from a recent selection of the most luminous $z>7$ galaxies in the CANDELS fields 
\citep{Roberts-borsani2015}, including record-breaking detections at $z=7.73$ \citep{Oesch2015} 
and $z=8.68$ \citep{Zitrin2015}.  The \citet{Roberts-borsani2015}  (hereafter RB16) photometric sample 
includes a total of four galaxies, each very bright in WFC3/IR imaging (H$_{160}=25.0-25.3$) with very red 
{\it Spitzer}/IRAC [3.6]-[4.5] colors, indicating extremely large equivalent width [OIII]+H$\beta$ emission.
How Ly$\alpha$ emission is able to escape efficiently from these systems 
while being so strongly attenuated from most other early galaxies is unclear.   
One possibility is that the most luminous galaxies trace overdense regions which produce the 
largest ionized bubbles at any given redshift (e.g., \citealt{Barkana2004,Furlanetto2004}).  
 \citet{Zitrin2015}  speculated that the selection of galaxies with red IRAC colors  
may pick out systems with hard ionizing spectra which are able to create early ionized bubbles, 
enhancing the transmission of Ly$\alpha$ through the IGM.    
 
With the aim of improving our understanding of the factors which are most important in regulating the escape of 
Ly$\alpha$ at $z>7$, we have recently initiated a comprehensive spectroscopic 
survey of Ly$\alpha$ and UV metal emission lines in the full photometric sample of galaxies identified in \citet{Roberts-borsani2015}.  
Our goals  are twofold.  Firstly, we seek a complete census of the Ly$\alpha$ emission equivalent widths.    
Thus far, only two of the four RB16 galaxies have been spectroscopically confirmed.   A third system 
(EGS-zs8-2) was found to have a tentative Ly$\alpha$ emission feature at $z=7.47$ in 
\citet{Roberts-borsani2015}, and the fourth system (COSY-0237620370, hereafter COS-zs7-1) has yet to be observed.  
Second, we aim to use knowledge of the UV metal emission line properties to understand the evolving visibility of Ly$\alpha$ 
emission at $z>7$.   In particular, we wish to characterize the hardness of the ionizing spectrum, determining if the RB16 galaxies are 
similar to the nebular CIV emitting $z=7.045$ galaxy reported in \citet{Stark2015b}.   Using the systemic redshift 
provided by the [CIII], CIII]$\lambda\lambda$1907,1909 doublet, we will investigate the velocity offset of Ly$\alpha$ 
in the RB16 sample, one of the most important parameters governing the IGM attenuation provided to Ly$\alpha$ in the reionization era.

We adopt a $\Lambda$-dominated, flat Universe with $\Omega_{\Lambda}=0.7$, $\Omega_{M}=0.3$ and
$\rm{H_{0}}=70\,\rm{h_{70}}~{\rm km\,s}^{-1}\,{\rm Mpc}^{-1}$. All
magnitudes in this paper are quoted in the AB system \citep{Oke1983}.

\section{Keck/MOSFIRE Observations and Analysis}

We present new  observations of three of the four galaxies identified in \citet{Roberts-borsani2015}.  
Data were obtained over three separate  observing runs using the near-infrared multi-object spectrograph 
MOSFIRE \citep{Mclean2012} on the Keck I telescope.   
Details of the MOSFIRE observations are  summarized in Table 1.  The first observing run was 12-15 April 2015.
We observed EGS-zs8-2 in the Y-band, 
targeting the tentative Ly$\alpha$ detection reported in \citet{Roberts-borsani2015}.   
The seeing was between 0\farcs5 and 0\farcs8 and skies were clear in 4.0 hours of integration.  The integration time 
of individual Y-band exposures was 180 seconds.   
 Both EGS-zs8-1 and EGS-zs8-2 were then observed in the H-band to constrain the strength of the 
 [CIII],CIII]$\lambda\lambda$1907,1909 doublet.    Conditions were mostly clear and seeing was 0\farcs5 during the 
2.5 hours of on-source integration.   
On  11 June 2015, we obtained an additional 1.0 hr of H-band observations on EGS-zs8-1 and EGS-zs8-2 
 in clear conditions with average seeing of 0\farcs6, bringing the total H-band integration time to 3.5 hrs.   The 
 individual H-band exposures are 120 seconds.
Finally, on 30 November 2015, we obtained a Y-band spectrum of COS-zs7-1.   
Conditions were clear and the average seeing was  0\farcs7.   We obtained 48 exposures of 180 seconds, totaling 
2.4 hours of on-source Y-band integration on COS-zs7-1.   Each mask contained 1-2 isolated stars for absolute 
flux calibration and numerous lower redshift galaxies.    We used slit widths of 0\farcs7 on all three observing runs.  

The publicly-available MOSFIRE Data Reduction Pipeline (DRP)\footnote{https://keck-datareductionpipelines.github.io/MosfireDRP/} was used to reduce the spectra.   The DRP 
performs flat-fielding, wavelength calibration, sky-subtraction, and cosmic ray removal, outputting 
reduced two dimensional spectra. For our chosen slit width,  MOSFIRE provides a resolving power of 
$R\simeq 3388$ (Y-band) and $R\simeq 3660$ (H-band), delivering a  
resolution of 2.82 ~\AA\ and 2.74 ~\AA\ in Y and H-band, respectively, for slit widths of 0\farcs7.     
Using the output from the DRP, we 
calculated two-dimensional signal to noise maps for each object.  One-dimensional spectra were then obtained using a 
boxcar extraction with apertures matched to the object profile, typically in the range 6-8 pixels (1\farcs08 - 1\farcs44).
The flux calibration was performed in two stages.  We determined the telluric correction spectrum 
via longslit observations of a spectrophotometric standard star conducted prior to the observations.  
The resulting correction spectrum accounts for the effects of the atmosphere and the instrumental response.
The absolute flux scale is calculated using the spectra of stars that were placed on the mask.  Using the {\it HST} flux 
measurments of 
the stars, we calculate the average scaling factor necessary to convert the observed count rate into flux density.   
The absolute scaling factor is applied to the telluric correction spectrum, resulting in a wavelength-dependent 
flux calibration which we apply to the 1D-extracted spectra.  If the sources are more extended than the stars 
used for the flux calibration, a small aperture correction will be required to account for slit losses.  We calculate the 
aperture correction individually for each source by convolving the {\it HST} images with the ground-based seeing.   
Given the very small sizes of the $z>7$ galaxies, these corrections are found to be negligible for the sources 
considered in this paper.

\begin{table*}
\begin{tabular}{lllccccccc}
\hline 
Source  & z$_{\rm{Ly\alpha}}$ & z$_{\rm{phot}}$ & RA & DEC & Date of Observations &   H$_{160}$ &  Filters &  UV lines  targeted  & Ref \\  \hline  \hline
EGS-zs8-1 &  7.730 & 7.92$^{+0.36}_{-0.36}$ & 14:20:34.89 & +53:00:15.4 & 12-15 Apr 2015 &  25.0 & H & CIII]   & [1], [2]   \\ 
\ldots & \ldots&\ldots & \ldots&\ldots &    11 June 2015&\ldots & H & CIII] & [1], [2] \\
EGS-zs8-2 &   7.477 & 7.61$^{+0.26}_{-0.25}$  & 14:20:12.09  &  +53:00:27.0 & 12-15 Apr 2015 & 25.1 & Y, H  & Ly$\alpha$,   CIII] & [1] \\ 
\ldots & \ldots&\ldots & \ldots&\ldots &    11 June 2015 &\ldots & H & CIII] & [1] \\
  COS-zs7-1 &  7.154 &  7.14$^{+0.12}_{-0.12}$  &  10:00:23.76 &  +02:20:37.0  & 30 Nov 2015  &  25.1& Y & Ly$\alpha$  & [1] \\
   \end{tabular}
\caption{Galaxies targeted with  Keck/MOSFIRE spectroscopic observations. The final column provides the reference to the article where each galaxy was first discussed in the literature.   The photometric redshifts shown in column three are taken from the discovery papers.  References: [1]  
\citet{Roberts-borsani2015}; [2]  \citet{Oesch2015}  }
\end{table*}

\section{Results}

\subsection{EGS-zs8-1}

\begin{figure*}
\begin{center}
\includegraphics[width=0.96\textwidth]{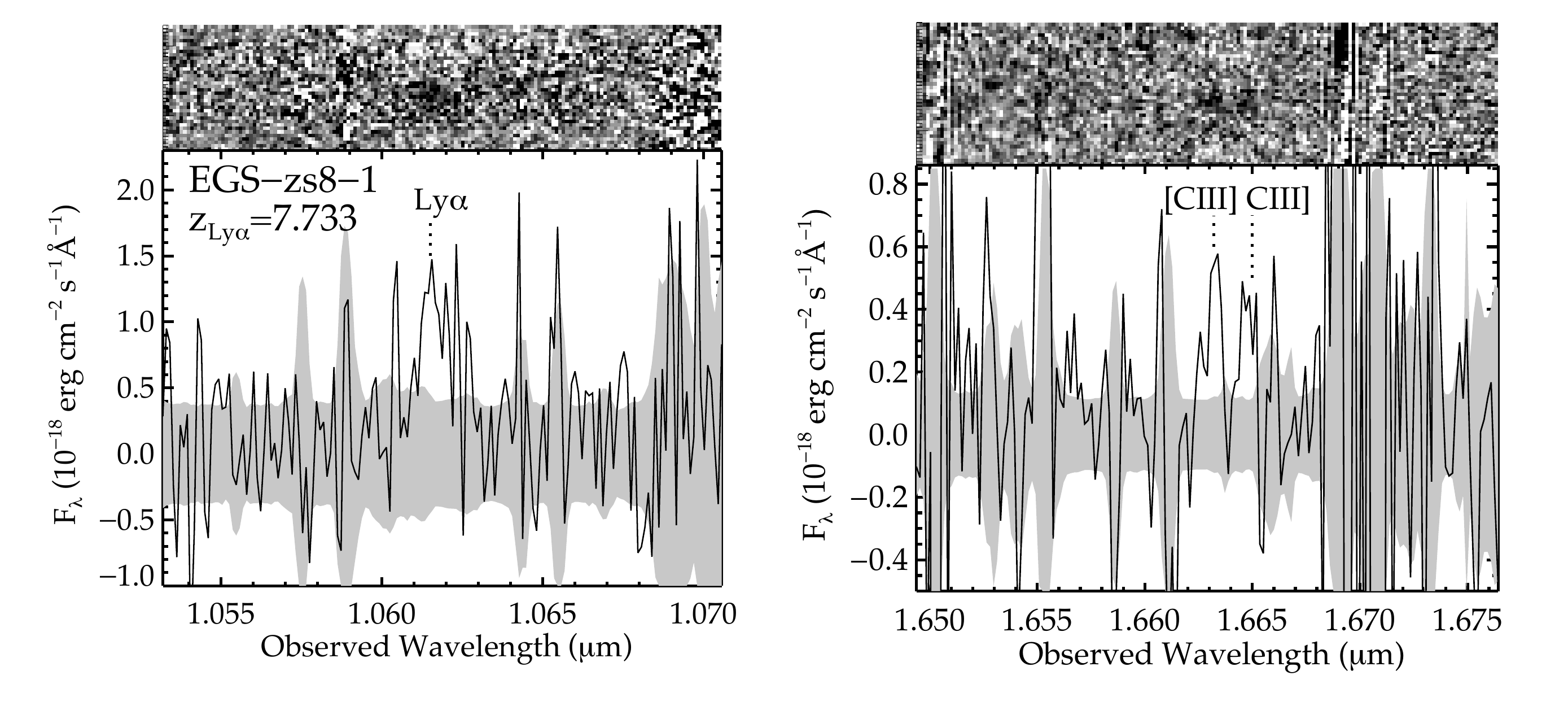}
\caption{Keck/MOSFIRE  spectra of EGS-zs8-1, a $z=7.733$ galaxy that was originally spectroscopically confirmed 
in \citet{Oesch2015}.  (Left:) Two-dimensional and one-dimensional Y-band spectra 
centered on the Ly$\alpha$ emission line.  Data are from \citet{Oesch2015}.  (Right:) 
H-band observations showing detection of the [CIII], CIII] $\lambda\lambda$1907,1909 doublet.
The top panels show the two dimensional SNR maps 
(black is positive), and the bottom panel shows the flux calibrated one-dimensional extractions.  }
\label{fig:egs-zs81}
\end{center}
\end{figure*}

EGS-zs8-1 is a bright (H$_{\rm{160}}$=25.0) galaxy with  a red IRAC color ([3.6]-[4.5]=$0.53\pm0.09$).   As described in \citet{Roberts-borsani2015}, the 4.5$\mu$m 
flux excess suggests very large rest-frame equivalent width ($911\pm122$\AA) [OIII]+H$\beta$ emission.  The optical line equivalent widths quoted here and 
below are inferred through computation of the line flux required to produce the measured IRAC flux excess, in the same manner as earlier studies 
\citep{Shim2011, Stark2013, Smit2014a,Smit2014b,Roberts-borsani2015},
The spectroscopic redshift was  confirmed by \citet{Oesch2015} through detection 
of a strong Ly$\alpha$ emission line using MOSFIRE (see Figure \ref{fig:egs-zs81}a).   The line 
reaches peak flux at 10616~\AA\, corresponding to $z_{\rm{Ly\alpha}}=7.733$.   \citet{Oesch2015} fit a truncated 
Gaussian profile to the data and calculate a redshift of $z_{\rm{Ly\alpha}}=7.730$.   
Since the true line profile is very uncertain, we will adopt the redshift set by the peak line flux for 
our analysis.  As we detail below, the result does not strongly impact our conclusions.
The Ly$\alpha$ line flux  (1.7$\times$10$^{-17}$ erg cm$^{-2}$ s$^{-1}$)  is one of the 
largest among $z>7$ galaxies and suggests a rest-frame equivalent width of W$_{\rm{Ly\alpha}}=21\pm4$~\AA. The 
absolute magnitude (M$_{\rm{UV}}=-22.1)$ indicates an unusually luminous galaxy, roughly 3$\times$ 
brighter than the value of L$^\star_{\rm{UV}}$ derived from the Schechter parameter fitting functions presented in \citet{Bouwens2015}.

The MOSFIRE H-band spectrum of EGS-zs8-1 covers the wavelength range between 1.5154 $\mu$m and 
1.8223~$\mu$m.   In Figure \ref{fig:egs-zs81}b, the narrow spectral window between 1.650 and 1.675$\mu$m
is shown, revealing a clean detection of both components of the [CIII], CIII] doublet.   We measure line 
fluxes of $4.5\pm0.5\times10^{-18}$ erg cm$^{-2}$ s$^{-1}$ and $3.6\pm0.5\times10^{-18}$ erg cm$^{-2}$ s$^{-1}$ 
for the 1907 and 1909 \AA\ components, respectively.   The total flux in the resolved CIII] doublet is close to 50\% that 
of Ly$\alpha$, nearly 10$\times$ greater than is seen in the most extreme CIII] emitting galaxies at lower redshift (e.g., \citealt{Erb2010,Christensen2012,Stark2014}).   
Since the continuum is undetected in the spectrum, we calculate the 
rest-frame equivalent widths using continuum flux derived from the broadband SED.    The measurements indicate a 
total [CIII], CIII] rest-frame equivalent width of $22\pm2$~\AA\ ($12\pm2$~\AA\ for [CIII]$\lambda$1907 and $10\pm1$~\AA\ for CIII]$\lambda$1909), 
similar to the value recently derived in a  gravitationally-lensed galaxy at $z=6.024$ \citep{Stark2015a}. 

The flux ratio of the [CIII],CIII] doublet provides a measurement of the electron density of the ionized gas.  
The ratio of  [CIII]$\lambda$1907/CIII]$\lambda$1909 varies from $\simeq 0.8$ for n$_e$=3$\times$10$^4$ cm$^{-3}$ 
to 1.5 for n$_e$=10$^2$ cm$^{-3}$.  
The measured [CIII]$\lambda$1907/CIII]1909 flux ratio of EGS-zs8-1 ($1.25\pm0.22$) suggests that CIII] traces reasonably high density gas.
The electron density of the system is determined by using IRAF's NEBULAR package (Shaw \& Dufour 1995).  
Assuming an electron temperature of 15,000 K,
consistent with metal poor CIII] emitting galaxies at lower redshifts (e.g., \citealt{Erb2010,Christensen2012}, Mainali et al. 2016 in prep), 
we infer an electron density of  9100$^{+12200}_{-7800}$ cm$^{-3}$ for EGS-zs8-1. The error bars on the measurement are calculated 
by including 1$\sigma$ error in the flux ratio as well as varying the electron temperature between 12,600 K  and 
20,000 K.   While uncertainties are clearly still significant, the CIII] density is noticeably larger than the average density (250 cm$^{-3}$) 
traced by [OII] or [SII] at $z\simeq 2.3$ \citep{Sanders2016}.   The tendency for CIII] to imply larger densities than [OII] and [SII] is well 
known (e.g., \citealt{James2014}, Mainali et al. 2016, in prep) and may indicate that the higher ionization line tends to be produced 
in denser regions within the galaxy.   Larger samples with multiple density diagnostics are required at lower redshift to assess whether the CIII] density offset 
is actually physical.  

The detection of CIII] also constrains the systemic redshift (e.g., \citealt{Erb2010,Stark2014}), providing a valuable
measure of the velocity offset of Ly$\alpha$, $\Delta v_{\rm{Ly\alpha}}$. 
The Ly$\alpha$ velocity offset is a key input parameter for models which seek to map the evolving 
number counts of Ly$\alpha$ emitters to IGM ionization state (e.g., \citealt{Choudhury2015}).  Yet owing to the 
lack of strong nebular lines in reionization-era galaxies, there are currently only a handful of $\Delta v_{\rm{Ly\alpha}}$ measurements at 
$z>6$ \citep{Willott2015,Stark2015a}. The wavelength centroids of the [CIII], CIII] doublet in EGS-zs8-1 reveal a systemic redshift of 7.723.   
The peak of the emergent Ly$\alpha$ emission line occurs at 10616 ~\AA, implying a velocity offset of 
$\Delta v_{\rm{Ly\alpha}}$=340$^{+15}_{-30}$ km s$^{-1}$.   The FWHM of the Ly$\alpha$ line (360 km s$^{-1}$; \citealt{Oesch2015}) 
thus indicates that a substantial fraction of the Ly$\alpha$ flux leaves the galaxy between  340 km s$^{-1}$ and 520 km s$^{-1}$. 
We note that if we were to adopt the truncated gaussian redshift for Ly$\alpha$, the inferred velocity offset would be somewhat 
smaller (260 km s$^{-1}$), but more importantly, the line profile still would imply a significant amount of flux emerging at yet larger velocities.  
The mean velocity offset of EGS-zs8-1 is comparable to the two measurements presented in \citet{Willott2015} but is 
considerably larger than that inferred in a robustly-detected CIII] emitter at 
$z=6.024$ \citep{Stark2015a} and well in excess of the $\Delta v_{\rm{Ly\alpha}}$ parameterization adopted in the 
reionization models of  \citet{Choudhury2015}.  We discuss the implications of these findings for reionization in \S5.2.

\begin{table*}
\begin{tabular}{lllccccccl}
\hline 
Source & z$_{Ly\alpha}$ & Line &  $\lambda_{\rm{rest}}$  &  $\lambda_{\rm{obs}}$  &  Line Flux  &  W$_0$ & H$_{\rm{160}}$ &  W$_{\rm{[OIII]+H\beta}}$& Ref \\ 
  &       & &  (\AA)                     &     (\AA)  & (10$^{-18}$  erg  cm$^{-2}$ s$^{-1}$) &    (\AA) & & (\AA) &   \\ \hline \hline
EGS-zs8-1 &  7.730 &  Ly$\alpha$   & 1215.67 &  10616 & $17\pm3$ & $21\pm4$  & 25.0 & $911\pm122$ & [1]\\ 
& &   $\rm{[CIII]}$  & 1906.68 & 16630  & $4.5\pm0.5$ & $12\pm2$ & \ldots &\ldots &This work  \\ 
 &  & $\rm{CIII]}$  & 1908.73 &   16645 & $3.6\pm0.5$ & $10\pm1$   & \ldots &\ldots  & This work \\ 
 EGS-zs8-2&  7.477  & Ly$\alpha$   & 1215.67 &  10305 & $7.4\pm1.0$ & $9.3\pm1.4$ & 25.1& $1610\pm302$ & [2], This work \\ 
   &&   $\rm{[CIII]}$  & 1906.68 & ---  & $<$2.3(3$\sigma$) & $<$7.1(3$\sigma$) & \ldots & \ldots& This work  \\ 
  &&  $\rm{CIII]}$  & 1908.73 &   --- & $<$2.3 (3$\sigma$)& $<$7.1(3$\sigma$) &  \ldots&  \ldots&This work \\ 
COS-zs7-1 &  7.154 &        Ly$\alpha$ & 1215.67 & 9913 & $25\pm4$ & $28\pm4$ & 25.1  & $1854\pm325$& This work\\  \hline \hline
EGS8p7 & 8.683 & Ly$\alpha$ & 1215.67 & 11774 & 17 &  28 & 25.3 &$895\pm112$ & [3] \\
z7\_GSD\_3811 & 7.664 &  Ly$\alpha$ &1215.67  &  10532 &  $5.5\pm0.9$ & 15.6$^{+5.9}_{-3.6}$ & 25.9 & --- & [4] \\
z8\_GND\_5296 & 7.508 &  Ly$\alpha$ &1215.67  &  10343 &  $2.6\pm0.8$ & $7.5\pm1.5$ & 25.6 & $1407\pm196$ & [5] \\
\ldots  &  7.508  & Ly$\alpha$   & 1215.67  &  10347 & $10.6\pm1.2$  & $46.9\pm5.4$  & \ldots  & \ldots  & [6] \\
SXDF-NB1006-2 & 7.215 & Ly$\alpha$ &1215.67  & 9988 & 19$^{+2.5}_{-0.9}$ & $>$15.4 & --- & --- & [7]   \\
GN-108036 & 7.213 & Ly$\alpha$ & 1215.67 & 9980 & 25 & 33 &25.2 (F140W) & $455\pm95$ & [8] \\
A1703\_zd6 &  7.045 & Ly$\alpha$ & 1215.67 & 9780 & $28.4\pm5.3$ &$65\pm12$  & 25.9 &--- & [9] \\
 & &  CIV & 1548.19 & 12458 & $4.1\pm0.6$ & $19.9\pm3.6$ & \ldots & --- & [10]  \\
BDF-3299 &  7.109 & Ly$\alpha$ & 1215.67 & 9858 &$12.1\pm1.4$ & 50 & 26.2 & --- & [11]  \\
BDF-521 & 7.008 &  Ly$\alpha$ &1215.67  & 9735 & $16.2\pm1.6$ & 64 &25.9 & --- &  [11]\\
 \end{tabular}
\caption{Rest-UV emission line properties of $z>7$ spectroscopically confirmed galaxies.   The top half of the table shows the targets observed in this paper.  
In the bottom half of the table, we include measurements for other sources in the literature with spectroscopic redshifts 
above $z\simeq 7$.  The equivalent widths include the aperture 
correction and are quoted in the 
rest-frame.   The upper limits are 3$\sigma$.  References: [1]  \citet{Oesch2015}; [2] \citet{Roberts-borsani2015}; [3] \citet{Zitrin2015}; [4] \citet{Song2016}; [5] \citet{Finkelstein2013};  [6] \citet{Tilvi2016}; [7] \citet{Shibuya2012}; [8] \citet{Ono2012};  [9] \citet{Schenker2012}; [10] \citet{Stark2015b}; [11] \citet{Vanzella2011}. }
\end{table*}

\subsection{EGS-zs8-2}
EGS-zs8-2 is another bright (H$_{\rm{160}}$=25.1) galaxy identified in CANDELS imaging by RB16.   The IRAC 
color of EGS-zs8-2 ([3.6]-[4.5]=$0.96\pm0.17$) is redder than EGS-zs8-1,  likely reflecting yet more extreme optical line emission.  We 
estimate a rest-frame [OIII]+H$\beta$ equivalent width of $1610\pm302$~\AA\ is required to reproduce 
the flux excess in the [4.5] filter.   A 4.7$\sigma$ emission feature was identified by \citet{Roberts-borsani2015} at 
a wavelength of 1.031$\mu$m.  RB16 tentatively interpret this feature as Ly$\alpha$.

\begin{figure*}
\begin{center}
\includegraphics[width=0.96\textwidth]{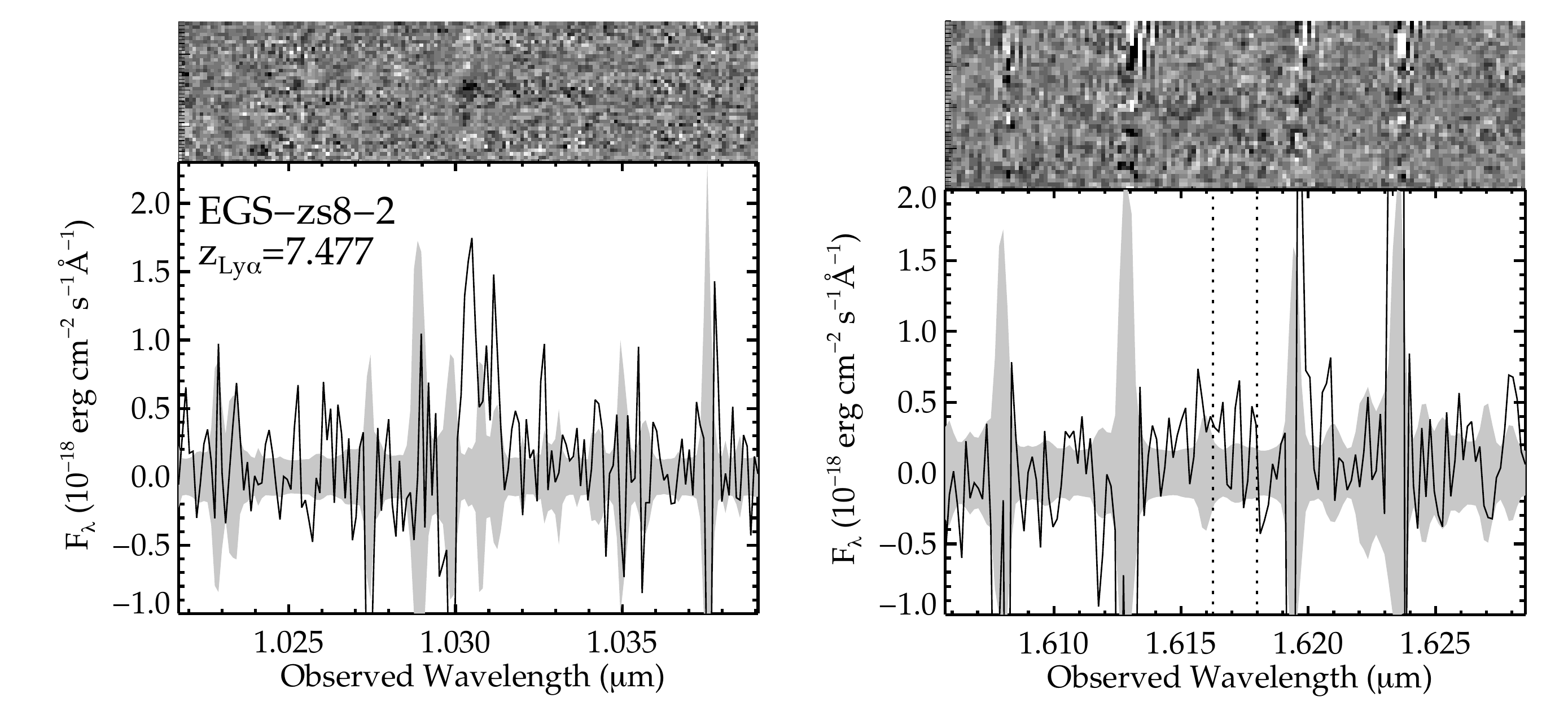}
\caption{Keck/MOSFIRE  spectra of EGS-zs8-2, a $z=7.477$ galaxy presented in RB16.  (Left:) Two-dimensional and one-dimensional Y-band spectra 
centered on the Ly$\alpha$ emission line, confirming the tentative redshift identification 
presented in \citet{Roberts-borsani2015}.  (Right:) 
H-band observations showing non-detection of the [CIII], CIII] $\lambda\lambda$1907,1909 doublet.
The top panels show the two dimensional SNR maps 
(black is positive), and the bottom panel shows the flux calibrated one-dimensional extractions.  }
\label{fig:egs-zs82}
\end{center}
\end{figure*}

We obtained a  Y-band spectrum of EGS-zs8-2 with the goal of verifying the putative Ly$\alpha$ detection.  The spectrum we 
obtained shows a 7.4$\sigma$ emission line at 1.0305 $\mu$m (Figure \ref{fig:egs-zs82}a), confirming that 
EGS-zs8-2 is indeed a Ly$\alpha$ emitter at z$_{\rm{Ly\alpha}}=7.477$.   The measured line flux ($7.4\pm1.0\times10^{-18}$ erg 
cm$^{-2}$ s$^{-1}$) is less than half that of EGS-zs8-1.   We calculate the Ly$\alpha$ equivalent width using the 
broadband SED to estimate the underlying continuum flux.   The resulting value (W$_{\rm{Ly\alpha}}$=$9.3\pm1.4$ \AA) is 
the smallest of the RB16 galaxies.  

The MOSFIRE H-band spectrum covers 14587 to 17914~\AA, corresponding to rest-frame wavelengths between 
1720 and 2113~\AA\ for EGS-zs8-2.   In Figure \ref{fig:egs-zs82}b, we show the spectral window centered on 
the [CIII], CIII] doublet.  No  emission lines are visible.   There are two weak sky lines in the wavelength range 
over which the doublet is situated.   However the separation of the individual components of the doublet is such that 
at least one of the two lines must be located in a clean region of the spectrum.   We estimate 3$\sigma$ upper 
limits of 2.3$ \times$10$^{-18}$ erg cm$^{-2}$ s$^{-1}$ for individual components.    The non-detection suggests 
that the total flux in the CIII] doublet must be less than 62\% of the observed Ly$\alpha$ flux, fully consistent with 
the ratio observed in EGS-zs8-1 and in extreme CIII] emitters at lower redshift.  We place a 3$\sigma$ upper limit on 
the doublet rest-frame equivalent width of $<$14~\AA.    Deeper data may yet detect 
CIII] in EGS-zs8-2.

\subsection{COS-zs7-1}

\begin{figure}
\begin{center}
\includegraphics[width=0.48\textwidth]{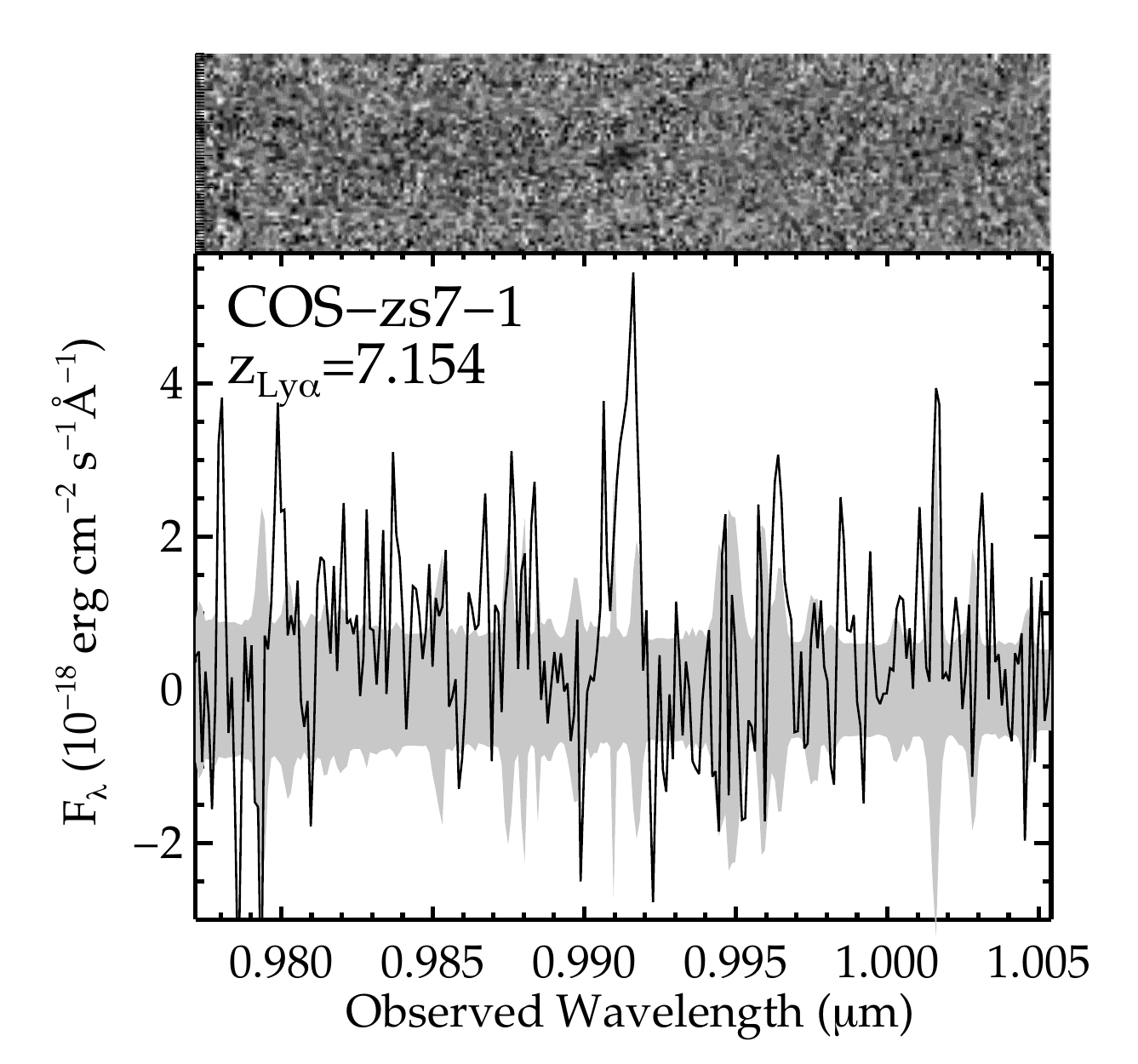}
\caption{Keck/MOSFIRE Y-band spectrum of COS-zs7-1, a bright (H=25.1) 
dropout presented in \citet{Roberts-borsani2015}.   We identify an emission feature 
at the spatial position of  the dropout at 9913~\AA\ which is likely to be Ly$\alpha$ at 
$z=7.154$.  The top panel shows the two dimensional SNR map 
(black is positive), clearly showing the characteristic negative-positive-negative signature 
expected from the subtraction of dithered data.   The bottom panel shows the flux calibrated one-dimensional extraction.   }
\label{fig:cos}
\end{center}
\end{figure}

Prior to this paper, COS-zs7-1 was the only source from \cite{Roberts-borsani2015} lacking   
a near-infrared spectrum.  Similar to the other galaxies from RB16, COS-zs7-1 is bright 
in the near-infrared (H$_{160}$=25.1) and has IRAC color ([3.6]-[4.5]=$1.03\pm0.15$) that indicates 
intense optical line emission.    In addition to RB16, the galaxy has been reported elsehwere (e.g., \citealt{Tilvi2013, Bowler2014}).
 We estimate an [OIII]+H$\beta$ rest-frame equivalent width of 
$1854\pm325$~\AA\ based on the [4.5] flux excess, making COS-zs7-1 the most extreme optical 
line emitter in the RB16 sample. RB16 derive a reasonably well-constrained photometric redshift 
(z$_{\rm{phot}}$=7.14$^{+0.12}_{-0.12}$) that  places Ly$\alpha$ in a narrow 290~\AA\ window between 9750 and 10041~\AA.

The Keck/MOSFIRE Y-band spectrum spans between 9750~\AA\ and 11238~\AA, covering the full range over which Ly$\alpha$ is predicted to lie.    We identify a 6.25$\sigma$ emission line at 9913~\AA\ that is  coincident 
with the expected spatial position of COS-zs7-1  (Figure \ref{fig:cos}).   The emission line is seen to have the 
standard negative - positive - negative pattern, indicating that it is present in both dither positions.
If the feature is Ly$\alpha$, it would correspond to  $z_{\rm{Ly\alpha}}=7.154$, in 
excellent agreement with the photometric redshift derived by RB16.   No other emission lines are visible at 
the spatial position of COS-zs7-1 in the Y-band spectrum.    We conclude that Ly$\alpha$ is the most likely 
interpretation of the line given the pronounced dropout in the $z$-band and the evidence for strong [OIII]+H$\beta$ 
emission in the [4.5] filter.  
The emission line is clearly distinct from sky lines with emission spanning 9905-9915~\AA; however the red side 
of the line coincides with positive residuals from a weak OH line at 9917~\AA, complicating the line 
flux measurement.  Integrating the emission line blueward of the OH line, we find a total flux of 
$2.5\pm0.4\times$10$^{-17}$ erg cm$^{-2}$ s$^{-1}$ and a rest-frame Ly$\alpha$ equivalent width of 
$28\pm4$~\AA. 

Table 2 summarizes the various emission line measures and the related physical properties
for all three sources in the context of earlier work.

\section{Photoionization Modeling}
The broadband SEDs of the RB16 galaxies suggest the presence of extremely large equivalent width [OIII]+H$\beta$ emission.  Here we  investigate whether the available data require an intense radiation field that may favor the escape of Ly$\alpha$.  In the case, of EGS-zs8-1 and EGS-zs8-2, we fold in the new constraints on [CIII], CIII] emission.   We focus our analysis on the three galaxies from RB16 for which we have obtained new spectral constraints (COS-zs7-1, EGS-zs8-1, EGS-zs8-2).
We fit the available  emission-line and broadband fluxes using the Bayesian spectral interpretation tool BEAGLE \citep{Chevallard2016}, which incorporates in a flexible and consistent way the production of radiation in galaxies and its transfer through the interstellar and intergalactic media. The version of BEAGLE used here relies on the models of Gutkin et al. (in preparation), who follow the prescription of \citet{charlot2001} to describe the emission from stars and the interstellar gas, based on a combination of the latest version of the \citet{Bruzual2003} stellar population synthesis model with the standard photoionization code CLOUDY \citep{ferland2013}. The main adjustable parameters of the photoionized gas are the interstellar metallicity, \zism, the typical ionization parameter of newly ionized \Hii\ regions, \Us\ (which characterizes the ratio of ionizing-photon to gas densities at the edge of the Stroemgren sphere), and the dust-to-metal mass ratio, \xid\ (which characterizes the depletion of metals on to dust grains). We consider here models with hydrogen density $\nH=100\,\mathrm{cm}^{-3}$, and two values of C/O abundance ratios, equal to 1.0 and 0.5 times the standard value in nearby galaxies [$(\mathrm{C/O})_\odot\approx0.44$.   Attenuation by dust is described using the 2-component model of \citet{charlot2000}, combined with the \citet{chevallard2013} `quasi-universal' prescription to account for the effects linked to dust/star geometry (including ISM clumpiness) and galaxy inclination. Finally, we adopt the prescription of \citet{inoue2014} to include absorption by the IGM.

We parametrize the star formation histories of model galaxies in BEAGLE as exponentially delayed functions $\psi(t)\,\propto\,t\,\exp(-t/\tauSFR)$, for star formation timescale in the range $7\leq\log(\tauSFR/\mathrm{yr})\leq10.5$ and formation redshift in the range $z_{\rm obs}\leq\zform\leq50$ (where $z_{\rm obs}$ is the observed galaxy redshift). We adopt a standard \citet{chabrier2003} initial mass function and assume that all stars in a given galaxy have the same metallicity, in the range $-2.2\leq\log(Z/Z_\odot)\leq0.25$. We superpose on this smooth exponential function a current burst with a fixed duration 10\,Myr, whose strength is parametrized in terms of the specific star formation rate, in the range $-14\leq\log(\psi_\mathrm{S}/\mathrm{yr}^{-1})\leq-7$. We adopt the same interstellar and stellar metallicity ($\zism=Z$) and let the dust-to-metal mass ratio and ionization parameter freely vary in the range
$\xid = 0.1-0.5$ and $-4\leq\log\Us\leq-1$, respectively. We consider $V$-band dust attenuation optical depths in the range $-3\leq\log\tauV\leq0.7$ and fix the fraction of this arising from dust in the diffuse ISM rather than in giant molecular clouds to $\mu=0.4$ \citep{wild2011}. 

With this parametrization, we use BEAGLE to fit the available constraints on the Ly$\alpha$ equivalent width (taken as a lower limit owing to resonant scattering), \ciiid\ equivalent width (for EGS-zs8-1 and EGS-zs8-2), and broadband F125W, F140W, F160W and IRAC 3.6\,$\mu$m and 4.5\,$\mu$m fluxes. We obtain as output the posterior probability distributions of the above free model parameters, as well as those of a large collection of derived parameters, such as for example the production rate of hydrogen ionizing photons per 1500\,\AA\ luminosity, $\xi^\ast_\textnormal{ion}$ (Table 3). The $\xi^\ast_\textnormal{ion}$ values correspond to the intrinsic UV emission from the stellar population model that reproduces the data, computed before reprocessing by gas and before attenuation by dust.  Below we also present $\xi_\textnormal{ion}$, which is computed considering the UV emission after it has been reprocessed by gas and attenuated by dust.   This latter quantity provides the total Lyman continuum production rate given the observed far-UV emission.
\begin{table*}
\begin{tabular}{|c|c|c|c|c|c|c|cc}
\hline
\hline
ID & $\log_{10}{[\textrm{sSFR}\ (\textrm{yr}^{-1})]}$ & $\tau_{V,\textrm{eff}}$ & $\log{U}$ & Z  & log$_{\rm{10}}$ $\xi^{\ast}_{ion}$ [erg$^{-1} \rm{Hz}$]   & $\xid$ &  [C/O]  \\
\hline 
EGS-zs8-1 &-7.66$_{-0.33}^{+0.26}$ & 0.01$_{-0.01}^{+0.02}$ & -1.61$_{-0.39}^{+0.37}$ & 1.7$_{-0.6}^{+0.8}\times10^{-3}$ & 25.59$_{-0.04}^{+0.03}$ & 0.28$_{-0.10}^{+0.10}$ & 0.52 \\
EGS-zs8-2  &-7.61$_{-0.39}^{+0.28}$ & 0.02$_{-0.02}^{+0.04}$ & -2.26$_{-0.46}^{+0.46}$ & 2.6$_{-1.5}^{+2.6}\times10^{-3}$ & 25.58$_{-0.04}^{+0.04}$ & 0.23$_{-0.09}^{+0.12}$ & 1.00 \\
COS-zs7-1 &-8.14$_{-0.29}^{+0.39}$ & 0.01$_{-0.01}^{+0.03}$ & -2.16$_{-0.48}^{+0.55}$ & 1.6$_{-0.7}^{+1.5}\times10^{-3}$ & 25.58$_{-0.09}^{+0.04}$ & 0.27$_{-0.11}^{+0.14}$ & 1.00 \\

\end{tabular}
\caption{Results from photoionization modeling using BEAGLE tool. The quoted uncertainties correspond to the 68\% central credible interval.  }
\end{table*}

The modeling procedure  is able to successfully reproduce the broadband SEDs of the 
\citet{Roberts-borsani2015} galaxies (i.e., Figure 4).  Matching the large flux 
excess in the IRAC [4.5] filter requires models with very large specific star formation rates 
(7-24 Gyr$^{-1}$), indicating a population undergoing rapid stellar mass growth.  
The implied interstellar metallicities are in the range 
Z=0.0016-0.0026, which is equivalent to 0.10-0.17 Z$_\odot$ using the solar metallicity 
value (Z$_\odot$=0.01524) from \citet{Bressan2012}.   
The strong [CIII], CIII] emission in EGS-zs8-1 forces the models to low metallicity (0.11 Z$_\odot$) 
and significantly reduces the allowable metallicity range.   Because of the depletion of metals onto dust grains (parameterized by the $\xi_d$ parameter) 
the gas-phase metallicity will be lower than the total interstellar metallicity that is fit by the models and reported in Table 3.  
 After accounting for the derived $\xi_d$ values using the method described in Gutkin et al. (2016, in prep), the 
gas-phase oxygen abundance is found to range between 12+log O/H = 7.76, 7.77, 7.97 for COS-zs7-1, EGS-zs8-1, EGS-zs8-2, respectively.   
The detection of 
[CIII], CIII]  in the spectrum of EGS-zs8-1 allows us to  consider variations in the C/O abundance ratio. 
We fit the broad-band fluxes and emission lines equivalent widths with the two different set of models corresponding to C/O$_\odot$, and 0.52 C/O$_\odot$. 
A visual analysis of the maximum-a-posteriori SED, and a comparison of the Bayesian evidence obtained with the two settings, indicates a slight 
preference of the model corresponding to 0.52 C/O$_\odot$, which exhibits a (marginally) larger sSFR, and a lower metallicity than the model with Solar-scaled C/O.  
The values reported in Table 3 thus correspond to the sub-Solar C/O models. 

As expected for galaxies dominated by such young and sub-solar stellar populations,  the models  suggest very large Lyman continuum photon 
production efficiencies,  log$_{\rm{10}}$ $\xi^{\ast}_{ion}$ [erg$^{-1} \rm{Hz}]  \simeq 25.6$, indicating that these 
galaxies have intense radiation fields.  The $\xi^{\ast}_{ion}$ values are  larger than canonical values commonly used in 
reionization calculations (e.g. \citealt{Kuhlen2012, Robertson2015,Bouwens2015c}), and are also larger 
than the average ionizing photon production efficiencies (log$_{\rm{10}}$  $\xi^{\ast}_{ion}$ [erg$^{-1}$ Hz] = 25.3) 
recently derived in  \citet{Bouwens2015d} for galaxies at $3.8<z<5.0$ (Figure \ref{fig:xi_ion}) and the average values derived at $z=2.2$ (log$_{\rm{10}}$  $\xi^{\ast}_{ion}$ [erg$^{-1}$ Hz]  = 24.77) by 
\citet{Matthee2016}.     If the extreme optical line emission of the RB16 galaxies 
is typical at $z>7$, it would indicate that reionization-era systems likely have considerably larger $\xi^{\ast}_{ion}$ values 
than previously thought, easing requirements on the escape fraction of ionizing radiation (e.g. \citealt{Robertson2015,Bouwens2015c}).   The estimated values of 
$\xi_{\mathrm{ion}}$ are in the range log$_{\rm{10}}$ $\xi_{\mathrm{ion}}$ [erg$^{-1}$ Hz]  = 25.60 and 25.73 for the three galaxies, larger than $\xi^{\ast}_{ion}$ 
because of the effect of dust attenuation which lowers the observed UV flux density.  

  \begin{figure}
\begin{center}
\includegraphics[width=0.48\textwidth]{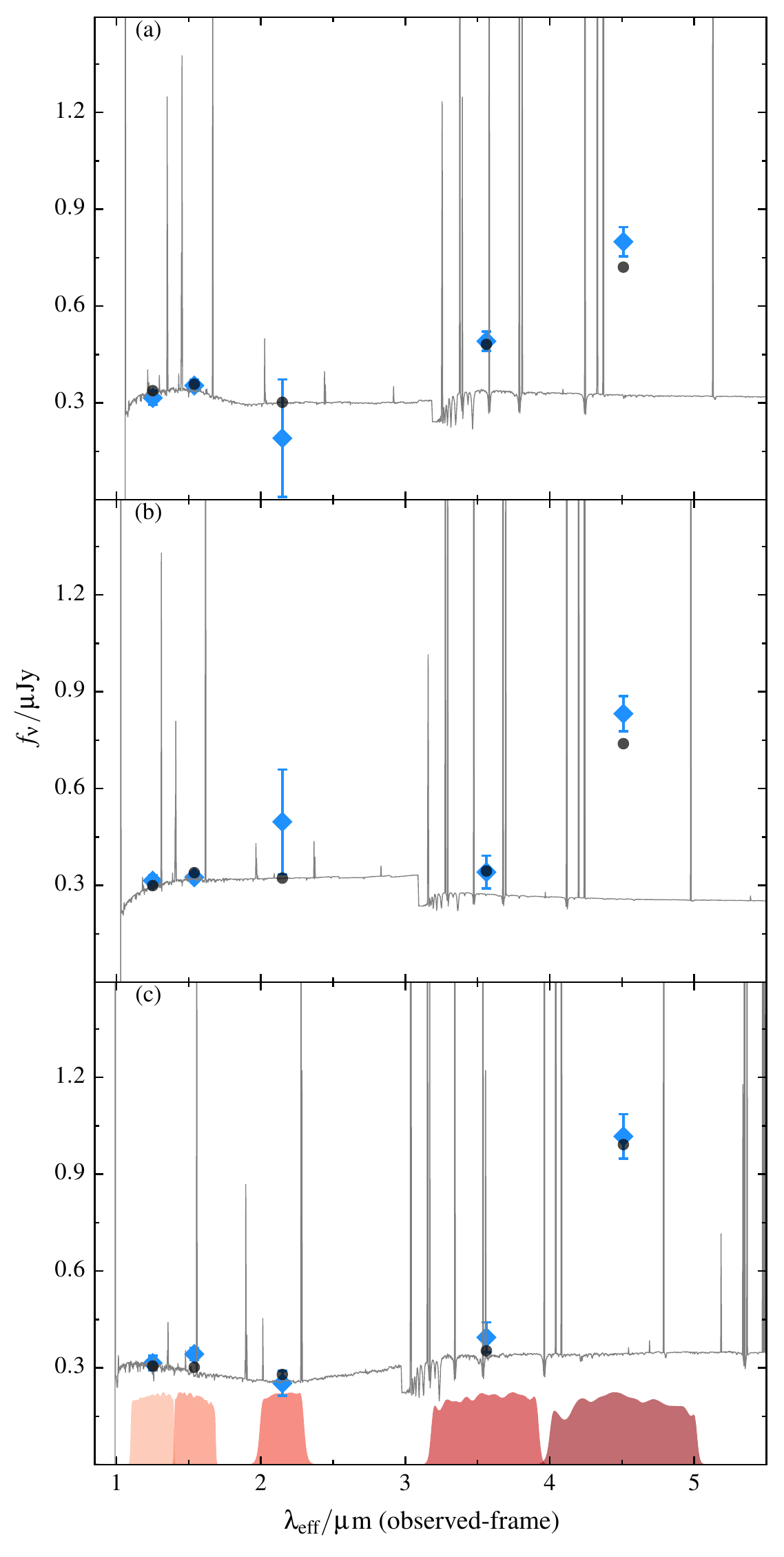}
\caption{Spectral energy distributions of EGS-zs8-1 (panel a), EGS-zs8-2 (panel b), and COS-zs7-1 (panel c).  The best-fitting BEAGLE 
SED models are overlaid.  Blue diamonds show the observed photometry reported in RB16.   The black circles show the synthetic photometry from BEAGLE.      }
\label{fig:seds}
\end{center}
\end{figure}
 \begin{figure}
\begin{center}
\includegraphics[width=0.48\textwidth]{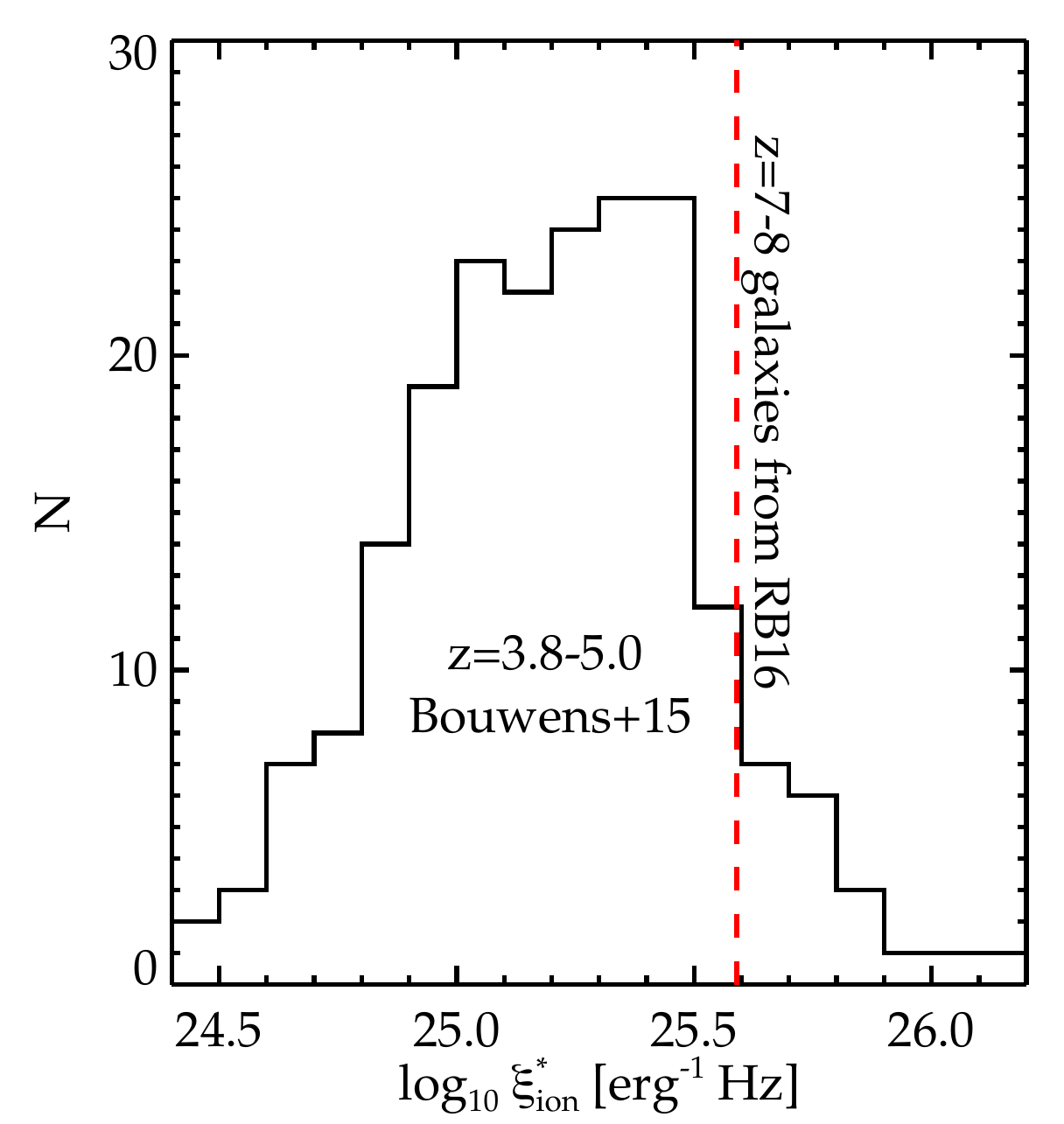}
\caption{A comparison of the Lyman continuum production efficiency, $\xi^{\ast}_{ion}$, for galaxies at $3.8<z<5.0$ (histogram) to EGS-zs8-1.   The selection of galaxies with IRAC color excesses picks out a population with high specific star formation rates 
and very large values of $\xi_{ion,unatt,\star}$. }
\label{fig:xi_ion}
\end{center}
\end{figure}

\section{High Fraction of Ly$\alpha$ Emission at $z>7$: A New Population?}

\begin{figure}
\begin{center}
\includegraphics[width=0.48\textwidth]{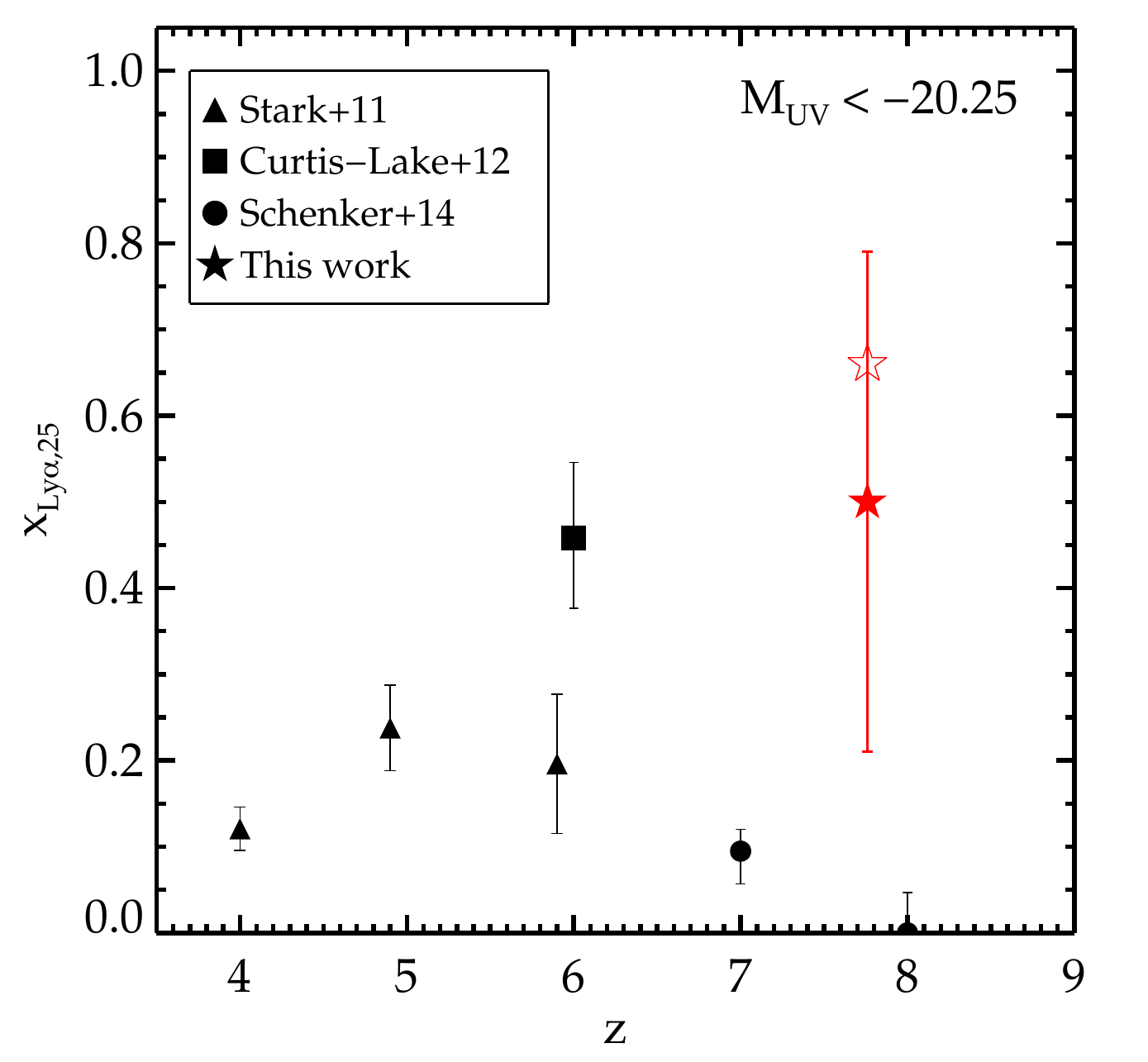}
\caption{The  fraction of Ly$\alpha$ emitters with W$_{\rm{Ly\alpha}} > 25$~\AA\ among UV luminous ($M_{\rm{UV}}<-20.25$) galaxies at $z>4$.   
The Ly$\alpha$ fraction in the RB16 sample is larger than found in previous studies of $z>7$ galaxies.  The open star shows the Ly$\alpha$ fraction that is derived 
using the new WFC3/IR grism measurement of W$_{\rm{Ly\alpha}}$ for the $z=7.508$ galaxy z8\_GND\_5296 \citep{Tilvi2016}, whereas the closed star shows 
the Ly$\alpha$ fraction derived using the MOSFIRE equivalent width measurement from  \citet{Finkelstein2013}.  
}
\label{fig:xlya}
\end{center}
\end{figure}

The detection of Ly$\alpha$ in COS-zs7-1 and EGS-zs8-2 establishes that all four of the galaxies identified  
in RB16 exhibit Ly$\alpha$ in emission, in spite of being situated at redshifts where the IGM is expected to be 
partially neutral.   Of the four Ly$\alpha$ emitters, two have W$_{\rm{Ly\alpha}} > 25$~\AA, implying a 
large Ly$\alpha$ fraction (x$_{\rm{Ly\alpha}}=0.50\pm0.35$ with W$_{\rm{Ly\alpha}}>25$~\AA).  We search 
previous spectroscopic studies (e.g., \citealt{Ono2012, Schenker2012,Finkelstein2013,Schenker2014,Pentericci2014}) 
for galaxies with photometric redshifts above $z\simeq 7$ that are both luminous (M$_{\rm{UV}}<-21$) and 
have IRAC fluxes indicative of intense [OIII]+H$\beta$ emission in the 4.5$\mu$m filter ($[3.6]-[4.5]>0.5$).   
Only two sources satisfy these requirements (most systems are either too faint or lack IRAC photometry), a $z=7.508$ 
galaxy reported in \citet{Finkelstein2013} and a $z=7.213$ 
galaxy confirmed in \citet{Ono2012}.  Both show Ly$\alpha$ emission, 
but only the $z=7.213$ emitter has a rest-frame equivalent width in excess of 25~\AA. \footnote{\citet{Tilvi2016} have recently presented detection of Ly$\alpha$ in the $z=7.508$ 
galaxy with the WFC3/IR grism as part of the FIGS survey (Malhotra et al. 2016, in prep).  The measured  Ly$\alpha$ flux (1.06$\times$10$^{-17}$ erg cm$^{-2}$ s$^{-1}$) and rest-frame equivalent width (46.9~\AA) are both 
larger than in the MOSFIRE discovery spectrum reported in \citet{Finkelstein2013}.   With the WFC3/IR grism measurement, the implied Ly$\alpha$ emitter fraction is larger yet: x$_{\rm{Ly\alpha}}=0.67\pm0.33$.}  Together with the RB16 sample, 
this implies a high Ly$\alpha$ fraction (x$_{\rm{Ly\alpha}}=0.50\pm0.29$ with W$_{\rm{Ly\alpha}}>25$~\AA).  
Although this conclusion appears at odds with previous studies at $7<z<8$ (Figure \ref{fig:xlya}),  
the average UV luminosity of the six galaxies with Ly$\alpha$ emission (M$_{\rm{UV}}=-21.9$) is larger than 
that of the galaxies in the luminous bin of the \citet{Schenker2014} measurements, possibly indicating that Ly$\alpha$ 
transmission may be enhanced in these ultra-luminous systems.  

The detection of UV metal lines allows us to begin exploring the precise physical mechanisms by which 
Ly$\alpha$  is able to escape so effectively from the luminous RB16 galaxies.   
 In \S5.1, we  consider whether the pre-selection of galaxies with  
IRAC [4.5] flux excesses is likely to influence the Ly$\alpha$ detection rate, and in \S5.2, we  use the systemic 
redshift provided by [CIII], CIII] to explore whether the Ly$\alpha$ velocity 
offsets of luminous galaxies boost the transmission of Ly$\alpha$ through the IGM.  We will argue that the selection of 
galaxies with IRAC color excess maximizes the production rate and transmission of Ly$\alpha$ through the local circumgalactic 
medium, while the identification of the brightest $z>7$ galaxies picks out sources which are most likely to transmit 
Ly$\alpha$ through the IGM.  

\subsection{Impact of local radiation field on Ly$\alpha$ equivalent widths}

The IRAC 4.5$\mu$m flux excesses of the RB16 sample are suggestive of extreme optical line emission.  
Photoionization models indicate that the data require very large specific star formation rates, moderately low 
metallicity, and large $\xi^{\ast}_{ion}$  (Table 3), suggesting efficient LyC production rates.   
The distribution of neutral hydrogen in the circumgalactic  medium could be very different in such 
young, rapidly growing systems.  Conceivably both the 
intense radiation field and enhanced stellar feedback of the RB16 galaxies could disrupt the surrounding distribution of gas, reducing the 
covering fraction of neutral hydrogen and boosting the transmission of Ly$\alpha$.   If the escape fraction of Ly$\alpha$ through the 
galaxy is indeed related to the specific star formation rate and $\xi^{\ast}_{ion}$, we should detect evidence of a 
larger Ly$\alpha$ emitter fraction in extreme optical line emitters located just after reionization ($4<z<6$).  

To test the connection between Ly$\alpha$ and $\xi^{\ast}_{ion}$, we investigate 
the Ly$\alpha$ equivalent width distribution in a large  
sample of $4<z<6$ galaxies described in our earlier work \citep{Stark2010,Stark2011,Stark2013}.    Redshifts 
were obtained via a large survey of UV-selected dropouts in GOODS-N using DEIMOS on Keck II (for details 
see \citealt{Stark2010}) and through a VLT/FORS survey described in by \citet{Vanzella2009}.   
Our goal is to determine whether  Ly$\alpha$ equivalent widths tend to be enhanced in the subset of $4<z<6$ galaxies 
with extreme optical line emission.   At $3.8<z<5.0$, it is possible to characterize rest-optical line emission using a 
similar IRAC flux excess technique as employed by RB16.   In this redshift range, the H$\alpha$ line is situated in the IRAC 
[3.6] filter, while the [4.5] band is free of strong emission lines (e.g. \citealt{Shim2011,Stark2013}).   While not identical 
to the RB16 selection (which identifies [OIII]+H$\beta$ emission), the subset of galaxies with extreme H$\alpha$ emission 
is similar in nature to those with extreme [OIII]+H$\beta$ emission \citep{Schenker2013b}.

In  \citet{Stark2013}, we measured H$\alpha$ equivalent widths for a sample of 
spectroscopically confirmed galaxies at $3.8<z<5.0$. Tang et al. (2016, in preparation) provide updated H$\alpha$ equivalent width 
measurements for the $3.8<z<5.0$  sample with spectroscopic constraints (including galaxies with and without redshift confirmations), 
making use of new high S/N near-infrared 
photometry from CANDELS and improved IRAC flux measurements, both extracted from catalogs 
provided in \citet{Skelton2014}.   Using this new catalog of $3.8<z<5.0$ H$\alpha$ measurements, we select those galaxies with IRAC [3.6] flux 
excesses indicative of H$\alpha$+[NII]+[SII] rest-frame equivalent widths in excess of 600~\AA.   This value is chosen by converting the 
[OIII]+H$\beta$ threshold of the \citet{Roberts-borsani2015} selection (W$_{\rm{[OIII]+H\beta}}$=900~\AA) to 
an H$\alpha$+[NII]+[SII] equivalent width using  the \citet{Anders2003} models with metallicity 0.2 Z$_\odot$.  
We identify thirty galaxies that satisfy this criterion.  In order to robustly compute a Ly$\alpha$ fraction, the sample 
includes galaxies regardless of whether we successfully confirmed a redshift.   To ensure the photometric subset is as 
reliable as possible, we follow previous studies (i.e. \citealt{Smit2016}) and only include those galaxies with photometric redshifts that are confidently 
within the $3.8<z<5.0$ redshift window.   Further details are included in Tang et al. (2016, in preparation).
The median UV luminosity and  spectroscopic(photometric) redshift of this subset of galaxies 
are M$_{\rm{UV}}=-20.6$  and 4.28(4.23), respectively.
We find that ten of thirty galaxies identified by this selection have W$_{\rm{Ly\alpha}}$ in excess of 25~\AA, 
implying a Ly$\alpha$ emitter fraction of x$_{\rm{Ly\alpha,25}}=0.33\pm 0.11$.   This is significantly greater than the Ly$\alpha$ emitter 
fraction of the full population of similarly luminous galaxies at $z\simeq 4$ 
($0.12\pm0.03$)  determined in \citet{Stark2011} and is significantly greater than that of galaxies with lower equivalent width H$\alpha$ emission 
(Tang et al. 2016, in preparation).   If we limit the extreme H$\alpha$ emitter sample to  galaxies with  UV continuum slopes that are similarly blue 
($\beta<-1.8$) as galaxies at $z>7$, we find an even larger Ly$\alpha$ emitter fraction (x$_{\rm{Ly\alpha,25}}=0.53\pm 0.17$).   
These results suggest that Ly$\alpha$ equivalent widths are boosted in galaxies with large equivalent width  
H$\alpha$ emission, reflecting either enhanced transmission or production of Ly$\alpha$ photons in galaxies with large sSFR.

A similar trend is seen in the faint  gravitationally-lensed $z\simeq 1.5-3$ galaxies described in \citet{Stark2014}.    
While the galaxies are much lower in stellar mass than the RB16 sample, they have similarly large specific 
star formation rates.   Five galaxies  have constraints on both [OIII]+H$\beta$ 
equivalent widths and Ly$\alpha$ emission.    The [OIII]+H$\beta$ equivalent widths of this subset are comparable to those in the RB16 
galaxies, ranging from 660~\AA\  to 1620~\AA.   The rest-frame Ly$\alpha$ equivalent widths of the five galaxies are also extremely large, 
ranging from 36 to 164~\AA, with a median of 73~\AA.

The results described above suggest that prominent Ly$\alpha$ emission (W$_{\rm{Ly\alpha}}>25$~\AA) is 
common in UV-selected galaxies with extreme optical line emission, suggesting a connection between the local 
radiation field and the escape of Ly$\alpha$ from the galaxy.   This could reflect both enhanced Ly$\alpha$ transmission 
(through the circumgalactic medium of the galaxy) and an unusually efficient Ly$\alpha$ production rate in systems with 
large specific star formation rates.  Based on these results, it is not surprising that a sample selected to have extreme 
optical line emission at $z>7$ is found to have larger-than-average Ly$\alpha$ equivalent widths at any given redshift.   
But unlike the $z\simeq 1.5-5$ galaxies described in this 
subsection, the RB16 systems are located at redshifts where the IGM is expected to be significantly neutral.  
While the transmission through the galaxy may be enhanced in the RB16 sample, it is not clear that 
the local radiation field is sufficient to boost transmission through the IGM.   One possibility, suggested in \citet{Zitrin2015}, is 
that the identification of the four most luminous $z>7$ galaxies in the CANDELS fields picks out overdense 
regions which have the largest ionized bubbles at any given redshift.    In the following subsection, we show 
that a correlation between Ly$\alpha$ velocity offsets and luminosity is likely to also contribute to the enhanced 
transmission of Ly$\alpha$ in the four RB16 galaxies.

\subsection{Impact of M$_{\rm{UV}}$ and $\Delta v_{\rm{Ly\alpha}}$ on Ly$\alpha$ transmission at $z>7$}

The velocity offset of  Ly$\alpha$ ($\Delta v_{\rm{Ly\alpha}}$)  plays an important role in 
regulating the transmission of the line through the IGM at $z>6$.  
The larger the  velocity offset from systemic, the further away Ly$\alpha$ will be 
from resonance by the time the line photons encounter intergalactic hydrogen.   The attenuation provided 
by the IGM to Ly$\alpha$ will thus be minimized in galaxies with large velocity offsets.  Knowledge 
of the typical velocity offsets of reionization-era galaxies is thus an important input for mapping the 
evolving Ly$\alpha$ counts to a neutral hydrogen fraction.

Unfortunately measurement of Ly$\alpha$ velocity offsets in the reionization era is extremely challenging. 
Not only is Ly$\alpha$ difficult to detect, but the standard rest-optical 
emission lines ([OIII], H$\alpha$) used to constrain the systemic redshift are not observable 
from the ground.  \citet{Erb2014} have recently characterized the Ly$\alpha$ velocity 
offsets for a large sample of galaxies at $z\simeq 2-3$, where [OIII] and H$\alpha$ can be 
easily detected with multi-object near-infrared spectrographs.   The results reveal 
several important relationships.   The velocity offset is correlated with UV luminosity and velocity dispersion 
(at $>$3$\sigma$ significance), and is anti-correlated with Ly$\alpha$ equivalent width 
(at 7$\sigma$ significance).   The smallest velocity offsets are thus found in low luminosity galaxies (Figure 7)
with small velocity dispersions and large Ly$\alpha$ equivalent widths. \citet{Erb2014} suggest a 
scenario where the Ly$\alpha$ profile is modulated by the properties of the gas at the systemic 
redshift.  The small values of $\Delta v_{\rm{Ly\alpha}}$ in low mass galaxies could reflect 
less developed gaseous disks (resulting in less neutral hydrogen at line center) and a harder 
radiation field capable of reducing the covering fraction of neutral gas at the systemic redshift.  

\begin{figure}
\begin{center}
\includegraphics[width=0.48\textwidth]{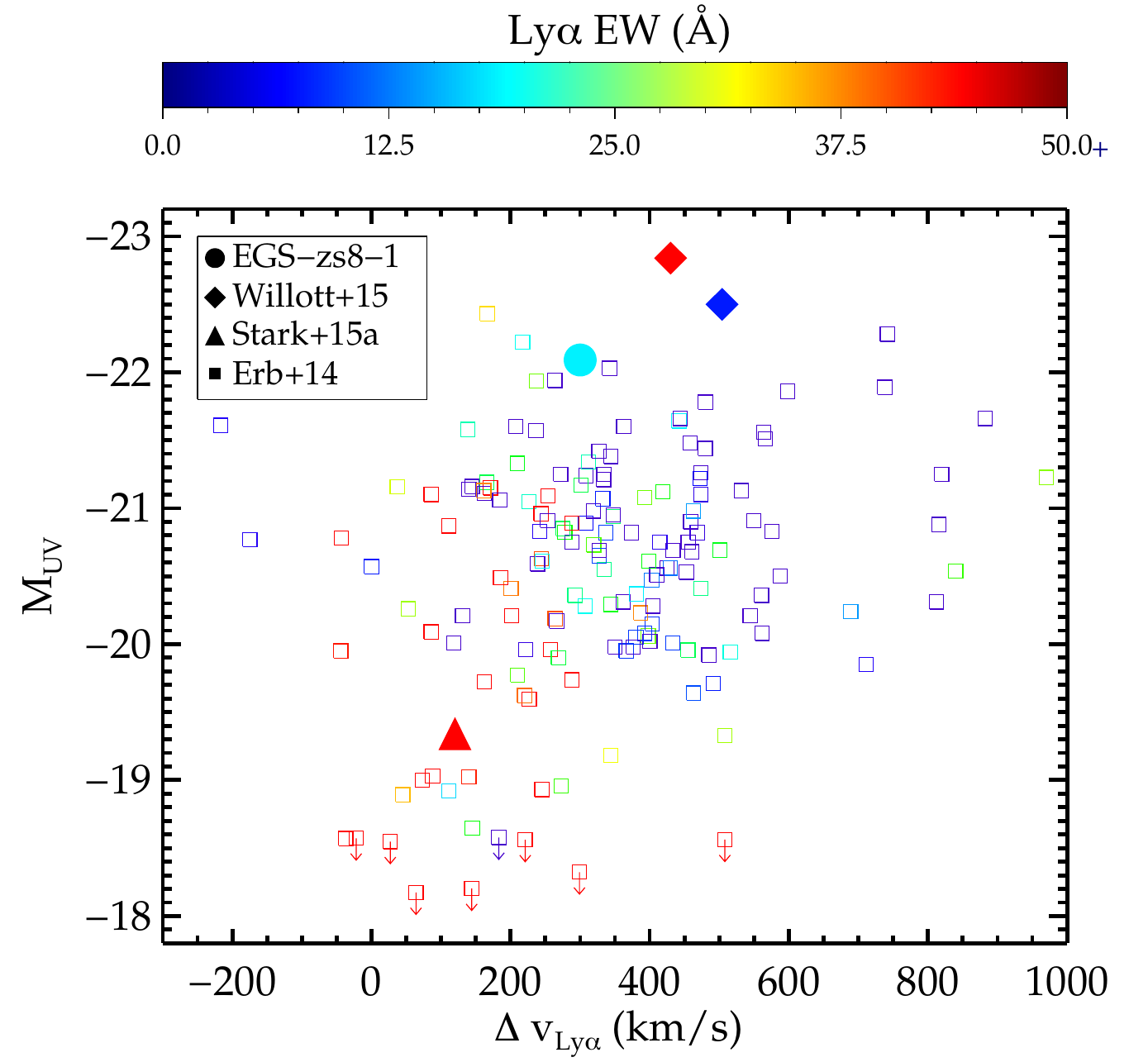}
\caption{The relationship between the Ly$\alpha$ velocity offset ($\Delta v_{\rm{Ly\alpha}}$) and the galaxy absolute UV magnitude 
(M$_{\rm{UV}}$).   The filled triangle, diamonds, and circle show existing constraints at $z>6$ from \citet{Stark2015a}, 
\citet{Willott2015}, and this study.   The  open squares show the $\Delta v_{\rm{Ly\alpha}}$-M$_{\rm{UV}}$ relationship derived at 
$z\simeq 2-3$ in \citet{Erb2014}.  The Ly$\alpha$ equivalent width is given by the color bar at the top of the plot, with red symbols corresponding to larger Ly$\alpha$ equivalent width than blue symbols.   }
\label{fig:dv_muv}
\end{center}
\end{figure}
 
The correlations described above provide a valuable baseline for predicting the likely range of Ly$\alpha$ 
velocity offsets expected in the reionization era.   Since galaxies at $z\simeq 6$ tend to have larger 
Ly$\alpha$ equivalent widths \citep{Ouchi2008, Stark2011, Curtislake2012, Cassata2015} and lower 
luminosities \citep{McLure2013,Bouwens2015,Finkelstein2015}, smaller Ly$\alpha$ velocity 
offsets are likely to be common in the reionization era.   Measurement of Ly$\alpha$ velocity offsets in a 
small sample of galaxies at $3.1<z<3.6$ \citep{Schenker2013b} provided the first  evidence of 
redshift evolution in  $\Delta v_{\rm{Ly\alpha}}$ at $z>2$ (Figure 8).   Most recently, \citet{Stark2015a} used the systemic 
redshift from detection of CIII]$\lambda$1909 in a a low luminosity (M$_{\rm{UV}}=-19.3$) 
$z=6.024$  galaxy to provide the first constraint on the $\Delta v_{\rm{Ly\alpha}}$ at $z>6$.  The 
measurement revealed a very small Ly$\alpha$ offset ($\Delta v_{\rm{Ly\alpha}}=120$ km s$^{-1}$), consistent 
with the trend reported in \citet{Schenker2013b}.   Such evolution could be driven by the emergence of 
harder radiation fields and the gradual disappearance of ordered gaseous disks, both of which would 
reduce the neutral hydrogen content at the systemic redshift.   If the tentative indications of evolution
in $\Delta v_{\rm{Ly\alpha}}$ are confirmed with future observations, it would require less intergalactic 
hydrogen to explain the disappearance of Ly$\alpha$ emitters at $z>6$  \citep{Mesinger2014}.
In particular,  \citet{Choudhury2015} have recently shown that the evolving Ly$\alpha$ fraction constraints 
can be fit with neutral fractions of $\simeq 30(50)$\% at $z\simeq 7$(8) if the average Ly$\alpha$ velocity 
offset decreases as (1+z)$^{-3}$ at $z>6$.

The connection between $\Delta v_{\rm{Ly\alpha}}$  and IGM transmission is likely to be very different 
for the luminous RB16 sample.   If the correlations discovered in \citet{Erb2014} are already in place 
at $z\simeq 7-8$, the velocity offsets will be larger, boosting the transmission of Ly$\alpha$ 
through the IGM.   The discovery of a 340 km s$^{-1}$   velocity offset 
in EGS-zs8-1 (see \S3.1) is consistent with this framework (Figure 7), 
suggesting that the most luminous galaxies at $z>7$ may have enough neutral gas at their systemic 
redshift to modulate the Ly$\alpha$ profile.    Additional support for the existence of a relationship between M$_{\rm{UV}}$ and 
$\Delta v_{\rm{Ly\alpha}}$ in the reionization era comes from the  discovery of  
large velocity offsets ($\Delta v_{\rm{Ly\alpha}}$ = 430, 504 km s$^{-1}$) in two of the most luminous galaxies known at 
$z\simeq 6$ \citep{Willott2015}.     Further data are clearly required to determine the relationship between 
$\Delta v_{\rm{Ly\alpha}}$ and M$_{\rm{UV}}$ at $z\simeq 6$, yet the first results suggest a scenario whereby 
large velocity offsets of luminous galaxies allow Ly$\alpha$ to be more easily transmitted through the surrounding 
IGM.   Since luminous systems are also likely to trace overdense regions within large ionized bubbles, the likelihood 
of detecting Ly$\alpha$ should be considerably greater in the most luminous galaxies at $z>6$.    Evidence for 
luminosity-dependent evolution of LAEs at $z>6$ has been suggested in previous Ly$\alpha$ fraction studies 
(e.g., \citealt{Ono2012}) and is also consistent with the lack of evolution at the bright end of the Ly$\alpha$ luminosity function over $5.7<z<6.6$
(e.g., \citealt{Matthee2015}).

\begin{figure}
\begin{center}
\includegraphics[width=0.48\textwidth]{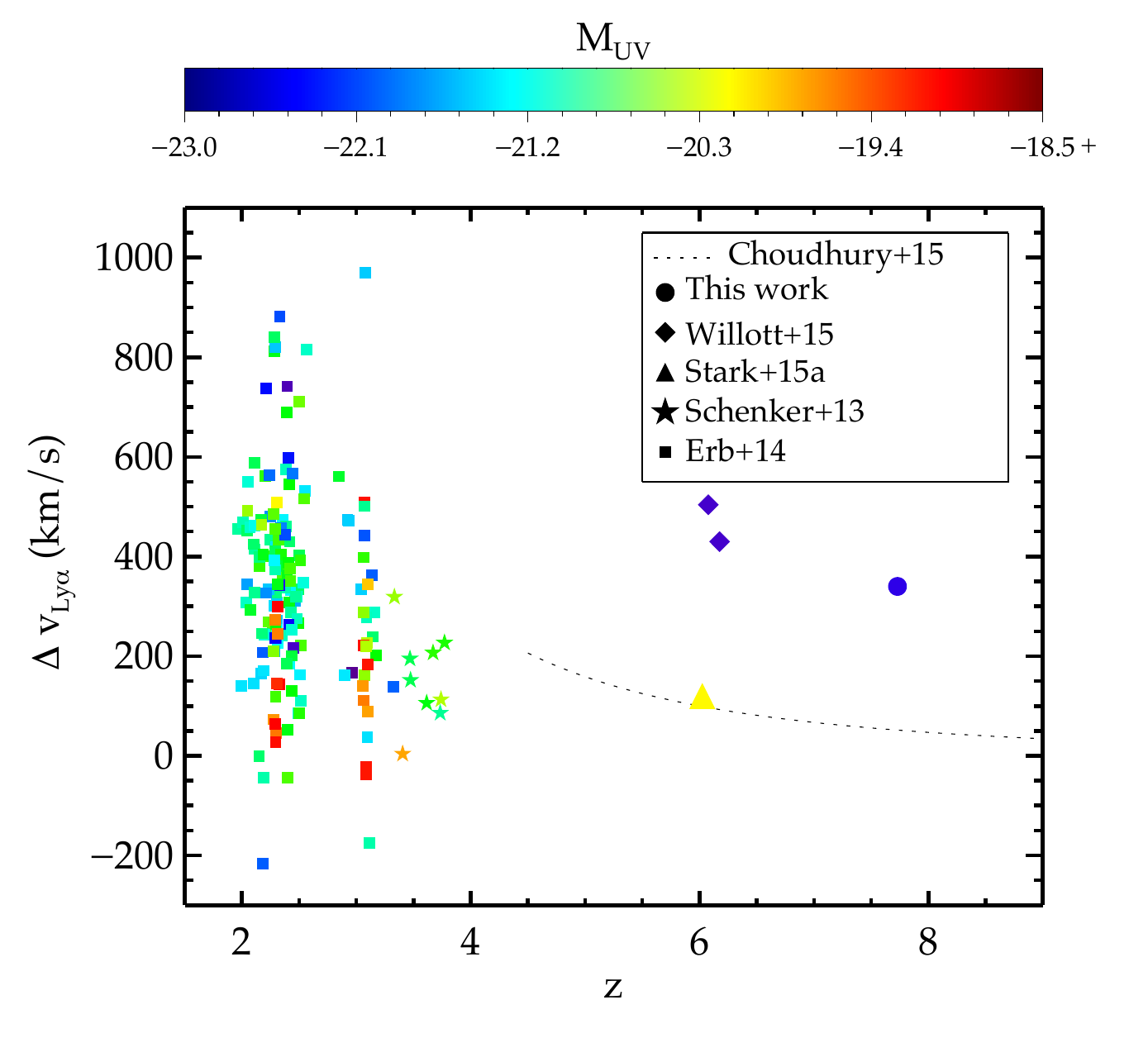}
\caption{ The relationship between the Ly$\alpha$ velocity offset  ($\Delta v_{\rm{Ly\alpha}}$) and redshift. 
The top color bar indicates the absolute magnitude, M$_{\rm{UV}}$, of individual galaxies.  The model of 
velocity offsets adopted in \citet{Choudhury2015} is shown as the dotted line.   Ly$\alpha$ observations at $z>6$ are likely to be biased 
toward detection of systems with large Ly$\alpha$ velocity offsets.}
\label{fig:dv_z}
\end{center}
\end{figure}

\section{Summary}

We present new Keck/MOSFIRE spectroscopic observations of  three of the four luminous $z>7$ galaxies presented in 
\citet{Roberts-borsani2015}.   The galaxies are selected to have a large flux excess in the [4.5] IRAC filter, indicative 
of intense [OIII]+H$\beta$ emission.  Previous spectroscopic follow-up has revealed Ly$\alpha$ emission 
in two of the four galaxies and a tentative detection in a third system.  Our new MOSFIRE observations confirm that
Ly$\alpha$ is present in the entire sample.   We detect  Ly$\alpha$ emission in the galaxy COS-zs7-1, confirming its redshift as 
$z_{\rm{Ly\alpha}}=7.154$, and we detect Ly$\alpha$ in EGS-zs8-2 at $z_{\rm{Ly\alpha}}=7.477$, verifying the lower S/N 
detection presented in \citet{Roberts-borsani2015}.

The ubiquity of Ly$\alpha$ emission in this photometric sample is puzzling given that the IGM is expected to be 
significantly neutral over $7<z<9$.   To investigate the potential implications for reionization, we have initiated a 
campaign to target UV metal line emission in the four Ly$\alpha$ emitters.  We present the detection of very 
large equivalent width [CIII], CIII] $\lambda\lambda$1907,1909 \AA\ emission in EGS-zs8-1 (W$_{\rm{CIII],0}}=22\pm2$~\AA), 
a galaxy previously shown by \citet{Oesch2015} to have Ly$\alpha$ emission at $z=7.73$.   The centroid of CIII] reveals 
that Ly$\alpha$ is redshifted from systemic by 340$^{+15}_{-30}$ km/s. 
This velocity offset is larger than that commonly found in less luminous systems and suggests that 
a correlation between velocity offset and luminosity, known to exist at $z\simeq 2$ \citep{Erb2014}, may already be in 
place in the reionization era.  Physically, the velocity offset is modulated by the properties of neutral hydrogen at the systemic 
redshift of the galaxy.    The existence of large velocity offsets at $z>6$ suggests that a 
substantial amount of gas has already accumulated at the line center in the most massive galaxies, forcing Ly$\alpha$ to escape 
at redder wavelengths.  We consider the requirements to match the broadband SEDs and UV metal line properties of the \citet{Roberts-borsani2015} 
galaxies using the new BEAGLE tool \citep{Chevallard2016}.   The red IRAC colors require the presence of  
an hard ionizing spectrum (log$_{\rm{10}}$ $\xi^{\ast}_{\rm{ion}}\simeq 25.6$) in all of the galaxies, while the detection of 
intense [CIII], CIII] emission in EGS-zs8-1  additionally requires models with reasonably low 
metallicity (0.11 Z$_\odot$).   

These initial results provide the context for understanding why Ly$\alpha$ appears so frequently in the luminous 
sample of galaxies discovered in \citet{Roberts-borsani2015}.   
The observability of Ly$\alpha$ at $z>7$ depends on the transmission through  
both the galaxy and the IGM.   We argue that the product of both quantities is maximized in the RB16 sample.   
The pre-selection of galaxies with extremely large equivalent width [OIII]+H$\beta$ emission picks out 
systems with very massive, young stellar populations.  Based on results at lower redshift, we suggest that the 
hard radiation field of these galaxies likely increases the production rate of Ly$\alpha$ and 
may also decrease the covering fraction of neutral hydrogen in the circumgalactic medium, boosting the transmission 
of Ly$\alpha$ through the galaxy.  Howewer, unlike at lower redshifts, the Ly$\alpha$ 
emission produced by the RB16 galaxies  must traverse a partially neutral IGM.  The     
correlation between the Ly$\alpha$ velocity offset and luminosity offers the explanation for why Ly$\alpha$ is able to 
escape so effectively through the IGM from this luminous population of $z>7$ galaxies.   For the typical low luminosity systems 
in the reionization era, Ly$\alpha$ emerges 
close to line center and is thus strongly attenuated by the IGM.   But for the most luminous systems with substantial 
velocity offsets, Ly$\alpha$ will be redshifted further into the damping wing by the time it encounters  intergalactic HI, 
enhancing the transmission through the IGM. The escape of Ly$\alpha$ will be  further amplified if luminous 
systems trace overdense regions situated in large ionized bubbles (e.g., \citealt{Furlanetto2004}).  As a result, 
the disappearance of the Ly$\alpha$ emitter population may well be less pronounced in the most luminous 
(i.e., M$_{\rm{UV}}=-22$) galaxies in the reionization era.   

\section*{Acknowledgments}
We thank Mark Dijkstra, Dawn Erb, and Martin Haehnelt for enlightening conversations.   We are grateful to  
Dawn Erb  for providing data on $z\simeq 2-3$ Ly$\alpha$ velocity offsets.
DPS acknowledges support from the 
National Science Foundation  through the grant AST-1410155.  RSE acknowledges
support from the European Research Council through an Advanced Grant FP7/669253.
SC, JG and AVG acknowledge support from the ERC via an Advanced Grant under grant agreement no. 321323 -- NEOGAL
AZ is supported by NASA through Hubble Fellowship grant \#HST-HF2-51334.001-A awarded by STScI, which is 
operated by the Association of Universities for Research in Astronomy, Inc. under NASA contract NAS~5-26555.
This work was partially supported  by a NASA Keck PI Data Award, administered by the 
NASA Exoplanet Science Institute. Data presented herein were obtained at the W. M. Keck Observatory 
from telescope time allocated to the National Aeronautics and Space Administration through the agency's 
scientific partnership with the California Institute of Technology and the University of California. 
The Observatory was made possible by the generous financial support of the W. M. Keck Foundation.
The authors acknowledge the very significant cultural role that the
summit of Mauna Kea has always had within the indigenous Hawaiian community.
We are most fortunate to have the opportunity to conduct observations from this mountain.

\bibliographystyle{mn2e}
\bibliography{references}

\label{lastpage}

\end{document}